\RequirePackage{fix-cm}
\documentclass[twocolumn,epjc3]{svjour3}  
\smartqed  
\usepackage{cite}
\usepackage{float}
\usepackage{amsmath}
\RequirePackage{graphicx}
\usepackage{hyperref}
\hypersetup{
   colorlinks=true,
    linkcolor=blue,
    filecolor=magenta,      
    urlcolor=blue,
    pdftitle={Sharelatex Example},
    bookmarks=true,
    citecolor=blue,
    pdfpagemode=FullScreen,
}

\journalname{Eur. Phys. J. A}
\begin{document}

\title{Mass-Spectroscopy of [$bb][\bar{b}\bar{b}$] and [$bq][\bar{b}\bar{q}$] tetraquark states in a diquark-antidiquark formalism
}


\author{Rohit Tiwari\thanksref{e1,addr1}
        \and
        D.P. Rathaud\thanksref{addr1} 
        \and
        Ajay Kumar Rai\thanksref{addr1}
}

\thankstext{e1}{e-mail: rohittiwari843@gmail.com}


\institute{Department of Physics, Sardar Vallabhbhai National Institute of Technology, Surat-395007, Gujarat, INDIA. \label{addr1}}

\date{Received: date / Accepted: date}

\maketitle
\begin{abstract}
In this article, we utilise the non-relativistic potential model to calculate the mass-spectra of all bottom [$bb][\bar{b}\bar{b}$] and heavy-light bottom [$bq][\bar{b}\bar{q}$] (q=u,d) tetraquark states in diquark-antidiquark approximation. The four-body problem is reduced  into two-body problems by numerically solving the $Schr\ddot{o}dinger$ equation using a cornell-inspired potential along with relativistic correction term. The splitting structure of the tetraquark spectrum is described using spin-dependent terms (spin-spin, spin-orbit, and tensor). We have successfully calculated and predicted the masses of bottom mesons, diquarks and tetraquarks. The masses of S and P-wave tetraquark states [$bb][\bar{b}\bar{b}$] and [$bq][\bar{b}\bar{q}$], respectively, are found to be between 18.7-19.4 GeV and 10.4-11.3 GeV, in which the masses of S-wave [$bb][\bar{b}\bar{b}$] states are less than the 2$\eta_{b}$, $\eta_{b}\Upsilon$, and 2$\Upsilon$ threshold. Additionally, we investigated the $Z_b(10610)$ and $Z_ b(10650)$ states in the current model and found that they are 150 MeV  below the $BB^{*}$ and $B^{*}B^{*}$ thresholds.

\end{abstract}
%
\noindent
{\bf Program Summary and Specifications}\\
\begin{small}
\noindent
{Program title:}\\
{Licensing provisions:}\\
{Programming language:}\\
{Repository and DOI:}\\
{Description of problem:}\\
{Method of solution:}\\
{Additional comments:}\\
\end{small}
\section{Introduction}
\label{intro}
%
In the past two decades, many exotic hadrons were experimentally found and theoretically predicted, which increase the interest in non-conventional hadron spectroscopy among researchers \cite{exotic1, exotic2, exotic3, mod2}.
Exotic hadrons are those which have quark combinations other than the conventional baryons (qqq) and mesons ($q\bar{q}$) \cite{Gellmann,Gyang}.
Some of these exotic states have quark contents such as $qq\bar{q}\bar{q}$, $q\bar{q}q\bar{q}$, and $qqqq\bar{q}$ etc., which are explained as multiquarks or loosely bound molecules  \cite{mod2}. In 2003, the observation of the exotic state X(3872) aka $\chi_{c1}(3872)$ in the decay process of $B^{\pm} \rightarrow K^{\pm} \pi^{+} \pi^{-} J/\psi$, renewed interest in exotic states, and as a consequence, many other exotic states are still being observed experimentally to this day \cite{belle,cdf,D0}. 
Among all the discovered non-conventional states some of the expected to have tetraquark structure \cite{tr1,tr2,tr3,pdg,silver}.
\\
\\
All the heavy tetraquarks $QQ\bar{Q}\bar{Q}$, which often contain a charm or bottom quarks, are of great interest to tetraquark researchers. Among the heavy tetraquarks, the recent discovery of all-charm tetraquarks X(6900) and the study of all-bottom tetraquarks $[bb][\bar{b}\bar{b}]$ have been important in understanding quark confinement inside the tetraquarks \cite{cccc,bbbb}.
In 2017, the CMS collaboration discovered $\Upsilon(1S)$ pair formation in pp collisions at $\sqrt{s} = 8$ TeV and an excess at 18.4 GeV in the $\Upsilon(1S) l^{-} l^{+}$ decay channel was proposed in a subsequent preliminary study \cite{cms,y1s} and whereas RHIC reported similar observation at 18.2 GeV in Cu+Au collisions \cite{rhic}. However, the LHCb collaboration hasn't found any evidence in the $\Upsilon$(1S) $\mu^{-}$ $\mu^{+}$ invariant mass spectrum \cite{bbbb}. Hence, fully bottom tetraquark searches required intensive efforts at the experimental frontier.
\\
\\
%
While the double-heavy tetraquark sector sought the observation of di-mesonic structure with ingredients $bb\bar{u}\bar{d}$ in 1988 \cite{bbud}. In Refs.\cite{mass,1plus}, a narrow tetraquark structure $bb\bar{u}\bar{d}$ was predicted with spin-parity $J^{P}=1^{+}$ and mass 10389 $\pm$ 12 MeV.
\\
\\
The Belle collaboration reported two charged bottomonium-like resonances $Z_{b}(10610)$ and $Z_{b}(10650)$,(hence referred to as $Z_{b}$ and $Z'_{b}$) in the invariant mass distributions $e^{+}e^{-} \rightarrow \Upsilon(nS)\pi^{+}\pi^{-}$, $n=1,2,3$ and $e^{+}e^{-} \rightarrow h_{b}(mP)\pi^{+}\pi^{-}$, $m=1,2$ \cite{zb}. Both $Z_{b}$ and $Z'_{b}$ were predicted in the decays of vector bottomonium $\Upsilon(10860)$ and afterwards verified in the elastic $B^{*}\bar{B}^{*}$ channels \cite{zb1}. The decays of the bottomonium-like tetraquarks (bound diquarks–
antidiquarks) into conventional bottomonium states and a pion indicate that these resonances have a minimum quark content of four quarks \cite{zb2,aslam,aslam2,zb4,zb5}. The study of double bottom tetraquark states have been attempted in various theoretical approaches like a quark model formalism \cite{life}, relativistic quark model \cite{nr}, and quark coalescence model \cite{lhc}. 
The study of all bottom tetraquark states have been attempted in various approaches like non-relativistic effective field theory \cite{nre}, QCD sum rule \cite{qcdsum}, the diquark-antiquark formalism \cite{dianti}, diffusion Monte Carlo approach \cite{monte}, an effective potential model \cite{QQQQ} etc. and predicted mass in the range of 18-18.8 GeV \cite{18g}.
\\
\\
Morover, the antiparticle of tetraquark $\bar{b}\bar{b}ud$ with the same quantum number $J^{P}=1^{+}$, which has a mass of 10476 $\pm$ 24 $\pm$ 10 MeV and is stable against its decay process as determined by Lattice QCD \cite{anti}. Other studies using the same methodology \cite{other1,other2} also support the existence of this firmly confined tetraquark states. 
\\
\\
\textbf{Indeed, under the diquark approximation the $bb\bar{u}\bar{d}$ and $b\bar{b}u\bar{d}$ systems are equivalent but they vary significantly when considered on a full color basis, in which the $bb\bar{u}\bar{d}$ state is deeply bound, while the $b\bar{b}u\bar{d}$ state is clearly unbound \cite{life,arxiv}.}
\\
\\
The focus of this paper is on the all-bottom tetraquak $[bb][\bar{b}\bar{b}]$, which we shall refer to as $T_{4b}$, and the double bottom tetraquark $[bq][\bar{b}\bar{q}]$, (q=u,d) in a diquark-antidiquark configuration. We have utilized the non relativistic model to obtain the mass-spectra of $T_{4b}$ and $bq\bar{b}\bar{q}$ tetraquark states, by configuring four quark system ($QQ\bar{Q}\bar{Q}$) in two diquark [$QQ$]-antidiquark [$\bar{Q}\bar{Q}$] system. The diquark [$QQ$] and antidiquark [$\bar{Q}\bar{Q}$] are made up of two quarks (antiquarks) in antitriplet (triplet) color states. After producing the mass-spectra of heavy-heavy and heavy-light bottom mesons, the optimal set of parameters for tetra-quarks was fitted. The $Schr\ddot{o}dinger$ equation has been solved numerically using a Cornell-like potential model and the relativistic correction term $\mathcal{O}(\frac{1}{m})$. Additionally, we included spin-dependent terms (spin-spin, spin-orbit, and tensor) in order to analyse the splitting of orbital and radial excitations.
\\
\\
The following is the structure of the current work: The theoretical model and fitting procedures are presented in Sec. II, followed by an introduction. Sec. III contains mass spectra of mesons and tetraquarks, as well as comparisons and discussions based on our model. We concluded our work in Section IV.
\section{Theoretical Model}
\label{sec:1}
We begin by introducing a non-relativistic model \cite{QQQQ} for spectroscopic analysis of hadronic bound states comprised of heavy-heavy and heavy-light quarks in the diquark-antidiquark model \cite{debastiani}. We first describe the four body system $QQ\bar{Q}\bar{Q}$ as two-body diquark [$QQ$] and antidiquark [$\bar{Q}\bar{Q}$] system. The mass-spectra of $T_{4b}$ and $bq\bar{b}\bar{q}$ tetraquarks states have been calculated by solving the $Schr\ddot{o}dinger$ equation numerically using code originally developed by W. Lucha et al. \cite{lucha} which is based on the fourth-order Runge-Kutta (RK4) method. The suitable method is to work in the center-of-mass frame while solving two-body problems which are included in the central potential \cite{prd}. To separate the angular term and radial term of a wave function spherical harmonics can be used. In the case of quarkonium and tetraquarks, the kinetic energy of states may be represented in terms of the reduced mass $\mu = \frac{m_{1} m_{2}}{m_{1}+ m_{2}}$, where $m_{1}$ and $m_{2}$ are the masses of constituents.
\\
\\ 
In the spectroscopic study, of heavy quark system the kinetic energy is comparatively lower than the rest mass energy of the constituent quarks, hence the reasonable approximation could be using of static potentials in a non-relativistic model \cite{exotic2}. The spin-dependent terms are incorporated perturbatively in the potential model. This methodology generates four free optimal set of parameters which are fitted to mesons-spectra, later it can be used in diquarks and tetraquarks. The Hamiltonian may be represented with an unperturbed one-gluon exchange (OGE) potential \cite{mod2} and the relativistic mass-correction term $V^{1}(r_{ij})$. The simple two body hamiltonian in the center of mass frame of mesons and tetraquarks given as;
%
\begin{equation}
H=\sum_{i=1}^{2}(m_{i}+\frac{p_{i}^{2}}{2m_{i}})-T_{CM}+\sum_{j> i=1}^{2} V(r_{ij})
\end{equation}
Here $m_{i}$ is the constituent mass and $p_{i}$ is the relative momentum of the system, while $ V(r_{ij})$ is the interaction potential. $T_{CM}$ is the kinetic energy of the center-of-mass motion.
\\
The time-independent radial $Schr\ddot{o}$dinger \cite{scr1,scr2} equation for two body problem can be expressed as; 

\begin{eqnarray}
\left[-\frac{1}{2\mu}\left(\frac{d^{2}}{d(r_{ij})^{2}}+\frac{2}{r_{ij}}\frac{d}{d(r_{ij})}-\frac{L(L+1)}{r_{ij}^{2}}\right)+ V(r_{ij})
\right]\times \nonumber \\  
\psi(r_{ij}) = E\psi(r_{ij}) \quad \quad  
\end{eqnarray}

where, L and E are the orbital quantum number and energy eigenvalue respectively. By substituting $\psi(r_{ij}) = r_{ij}^{-1}\phi(r_{ij})$ in Eq.(1) modifies to;
\begin{equation}
\left[-\frac{1}{2\mu}\left(\frac{d^{2}}{dr_{ij}^{2}}+\frac{L(L+1)}{r_{ij}^{2}}\right)+V(r_{ij})\right]  \phi(r_{ij}) = E\phi(r_{ij})  
\end{equation}
The reliable and widely used potential model in the spectroscopic study of heavy-quarkonium system is a zeroth-order $V(r_{ij})$ Cornell-like potential \cite{cornell}. The potential $V_{C+L}(r_{ij})$ represents a gluonic interaction term between two (anti)quarks and (anti) di-quarks, referred to as the coulomb term, as well as a linear term responsible for quark confinement.
\begin{equation}
V_{C+L}(r_{ij})=\frac{k_{s}\alpha_{s}}{r_{ij}}+br_{ij}
\end{equation}
where, $\alpha_{s}$ is known as the QCD running coupling constant, $k_{s}$ is a color factor, b is string tension. We have incorporated the relativistic mass correction term $V^{1}(r_{ij})$ initially developed by Y. Coma et al. \cite{koma}, in the central potential. The final form of central potential is given by;
\begin{equation}
 V(r_{ij}) = V_{C+L}(r_{ij}) + V^{1}(r_{ij})\left( \frac{1}{m_{1}}+\frac{1}{m_{2}}\right)+\mathcal{O}\left(\frac{1}{m^{2}}\right)
\end{equation}
The non-perturbative form of relativistic mass correction term $V^{1}(r_{ij})$ is not yet known, but leading order perturbation theory yields \cite{koma},
\begin{equation}
V^{1}(r_{ij})=-\frac{C_{F}C_{A}}{4} \frac{\alpha^{2}_{s}}{(r_{ij})^{2}}
\end{equation}
where $C_{F}=\frac{4}{3}$ and $C_{A}=3$  are the Casimir charges of the fundamental and the adjoint representation respectively \cite{koma}.
The relativistic mass correction is found to be similar to the coulombic term of the static potential when applied to the charmonium and to be one-fourth of the coulombic term for bottomonium. Along with the central interaction potential $V(r_{ij})$, we have also incorporated spin-dependent interactions. These spin-dependent terms are included perturbatively.
\\
\\
\textbf{
The non-relativistic form of kinetic energy is given by $KE_{NR}=p^{2}/2m$ while semi-relativistic form is $KE_{SR}= \sqrt{p^{2}+m^{2}}-m$ \cite{relativistic,dpepjc}.}
In our previous studies \cite{ak,vk,dpepjc,dpepjp,dpijp1,dpfew,dpijp2}, we looked at the effect of relativistic correction of the kinetic energy portion for heavy-light mesons and di-mesonic molecule systems, and found that the contribution of higher order terms contributes less than $1\%$ to the net strength of the kinetic energy part. 
The proposed systems in this paper are considerably heavier, thus we think that higher order kinetic energy contribution should be extremely little and ignored in our calculations. \textbf{However, in Ref. \cite{relativistic}, a complete numerical study of the effects of relativistic corrections on the kinetic energy operator for $QQ\bar{q}\bar{q}$ and $QQ\bar{Q}\bar{Q}$ are presented.}
\subsection{Spin-dependent Terms}
The contribution of spin-dependent potentials, i.e. a spin-spin $V_{SS}(r_{ij})$, spin-orbit $V_{LS}(r_{ij})$, and tensor $V_{T}(r_{ij})$, that makes significant contributions particularly for excited states, is necessary to better understand the splitting between orbital and radial excitations of different combinations of quantum numbers of $T_{4b}$ and $bq\bar{b}\bar{q}$. All three spin-dependent terms are driven by the Breit-Fermi Hamiltonian for one-gluon exchange \cite{spin1,spin2}, and yields;

\begin{equation}
V_{SS} (r_{ij}) = C_{SS}(r_{ij})  S_{1} \cdot S_{2},
\end{equation}
\begin{equation}
V_{LS} (r_{ij}) = C_{LS}(r_{ij}) L \cdot S,
\end{equation}
\begin{equation}
V_{T} (r_{ij}) = C_{T}(r_{ij}) S_{12},
\end{equation}
The matrix element $S_{1} \cdot S_{2}$ acts on the wave function, and generates a constant factor, but the $V_{SS}(r_{ij})$ remains a function of only $r_ {ij}$, and the expectation values of $\left\langle S_{1} \cdot S_{2}\right\rangle$  are available through a quantum-mechanical formula \cite{griffiths}.

\begin{equation}
\left\langle S_{1} \cdot S_{2}\right\rangle =  \left\langle \frac{1}{2} (S^2 - S^2_{1}-S^2_{2})\right\rangle
\end{equation}
where, $S_{1}$ and $S_{2}$ denote the spins of constituent quarks for quarkonium and diquarks for tetraquarks, respectively. $C_{SS}(r_{ij})$ may be defined as follows:
\begin{equation}
C_{SS}(r_{ij}) = \frac{2}{3m^2} \nabla^2 V_{V}(r_{ij}) = -\frac{8k_{s}\alpha_{s}\pi}{3m^2} \delta^3(r_{ij}),
\end{equation} 
A fair agreement may be achieved by adding spin-spin interactions in a zero-order potentials using the $Schr\ddot{o}dinger$ equation in heavy quarkonium spectroscopy by including the spin-spin interaction using the artefact providing a new parameter $\sigma$ instead of the Dirac delta. So now $V_{SS}(r_{ij})$ can be redefined as;
\begin{equation}
V_{SS} (r_{ij}) = -\frac{8 \pi k_{s}\alpha_{s}}{3m^{2}} (\frac{\sigma}{\sqrt{\pi}})^{3} \exp^{-\sigma^2 (r_{ij})^2} S_{1} \cdot S_{2},
\end{equation} 
The expectation value of operator $\mathbf{\left\langle L \cdot S \right\rangle}$ is mainly dependent on the total angular momentum J which is calculated using the formula $\mathbf{J=L+S}$ ;
\begin{equation}
\mathbf{\left\langle L \cdot S \right\rangle =  \left\langle \frac{1}{2} (J^2 - L^2-S^2)\right\rangle}
\end{equation}
where, L denotes the total orbital angular momentum of quarks and diquarks, respectively, in the case of quarkonium and tetraquark. The following equation may be used to compute $C_{LS}(r_{ij})$ ;
\begin{equation}
C_{LS}(r_{ij}) = -\frac{3k_{s}\alpha_{s}\pi}{2m^2}\frac{1}{(r_{ij})^{2}}-\frac{b}{2m^2}\frac{1}{(r_{ij})}
\end{equation}
The second component in the spin-orbit interaction is called Thomas Precession, and it is proportional to the scalar term. It is thought that confining interaction originates from the Lorentz scalar structure. In higher excited states, the contribution of the spin-tensor becomes quite important, which requires a little algebra and may be calculated by;
\begin{equation}
V_{T}(r_{ij}) = C_{T}(r_{ij})\left( \frac{(S_{1}\cdot (r_{ij}))(S_{2}\cdot (r_{ij}))}{(r_{ij})^2}- \frac{1}{3} (S_{1} \cdot S_ {2})\right) 
\end{equation}
where;
\begin{equation}
C_{T}(r_{ij}) = -\frac{12k_{s}\alpha_{s}\pi}{4m^2}\frac{1}{(r_{ij})^{3}}
\end{equation}
The results of $(S_{1} \cdot S_ {2})$ may be obtained by solving the diagonal matrix elements for the spin-$\frac{1}{2}$ and spin-1 particles, as detailed in the following references\cite{debastiani,prd}. To solve the tensor interaction, the simpler formulation may be used ;
\begin{equation}
\mathbf{{S_{12}}}=12\left(\mathbf{ \frac{(S_{1}\cdot (r_{ij}))(S_{2}\cdot (r_{ij}))}{(r_{ij})^2}- \frac{1}{3} (S_{1} \cdot S_ {2})}\right) 
\end{equation} 
and which can be redefined as ;
\begin{equation}
\mathbf{{S_{12}}}= 4\left[3\mathbf{(S_{1}\cdot \hat{(r_{ij})})(S_{2}\cdot \hat{(r_{ij})})-(S_{1} \cdot S_ {2})} \right]
\end{equation}
Pauli matrices and spherical harmonics with their corresponding eigenvalues may be used to achieve the results of the $S_{12}$ term. The following conclusions are valid for bottomonium and diquarks \cite{bethe,thesis} ;
\begin{align}
\mathbf{\left\langle S_{12}\right\rangle}_{\frac{1}{2}\otimes\frac{1}{2}\rightarrow S=1, l\neq0} = -\frac{2l}{2l+3}, for J= l+1,\\
= -\frac{2(l+1)}{(2l-1)}, for J= l-1, and\\
= +2, for J=l
\end{align}
When $l=0$ and $S=0$ the $\mathbf{\left\langle S_{12}\right\rangle}$ always vanishes, but it yields a non-zero value for excited states in mesons: $\mathbf{\left\langle S_{12}\right\rangle} = -\frac{2}{5}, +2, -4$ for $J=2,1,0$, respectively.
These value are valid only for bottomonium and diquarks that are specifically spin-half particles, but in the case of tetraquarks when spin-1 diquarks are involved, it needs a laborious algebra, which is not discussed in depth here, rather one can refer Refs. \cite{bethe,thesis} for detailed discussion. The results for tensor interaction of $T_{4b}$ will obtained by the same formula which is used in case of bottomonium except that the wavefunction obtained here will be of spin 1 (anti)diquark.
\begin{align}
\mathbf{S_{{d}-{\bar{d}}}}=12\left(\mathbf{\frac{(S_{d}\cdot (r_{ij}))(S_{\bar{d}}\cdot (r_{ij}))}{(r_{ij})^2}- \frac{1}{3} (S_{d} \cdot S_ {\bar{d}})}\right)\\
= \mathbf{S_{14} + S_{13} + S_{24} + S_{23}}
\end{align}
where $S_{d}$ is the total spin of the diquark, $S_{\bar{d}}$ is the total spin of the antidiquark. When the two-body problem is solved to obtain the masses of the tetraquarks, the interaction between the two (anti) quarks inside the (anti) diquark is identical; because (anti) diquarks are only considered in the S-wave state, only the spin-spin interaction is relevant; the spin-orbit and tensor are both identically zero.
Because the tetraquark radial wavefunction is obtained by treating the diquark and antidiquark as two body problem, it is reasonable to assume that the radial-dependence of the tensor term is the same for these four [$q\bar{q}$] interactions and can be obtained using the radial wavefunction with Eq (14). The following functional form for spin $\frac{1}{2}$ particles does not use any specific relation or eigenvalues, instead relying on general angular momentum elementary theory \cite{tb}. Within this approximation, generalization of tensor operator can be consider a sum of four tensor interaction between four quark-antiquark pair as illustrated in \cite{thesis}.
A thorough discussion on tensor interaction can be found in Ref. \cite{thesis}.

\subsection{Fitting Parameters} 
We have obtained the mass-spectra of mesons and tetra-quarks for three set of fitting parameters. In the present work there are four fitting parameters (m, $\alpha_{s}$, b, $\sigma$) for which the model mass ($m_{i}^{f}$) of the particular tetraquark states have been calculated. 
\\
The best fitting parameters minimizes the difference between an experimental mass ($m_{i}^{exp}$) and mass ($m_{i}^{f}$) obtained from  model, which we will call discrepancy and denoted as $\Delta=[m_{i}^{exp}-m_{i}^{f}]$. The set of parameters \cite{pdg,griffiths} are varied in the range of ;

{\centering
0.05 $\leq$ $\alpha_{s}$ $\leq$ 0.70\\
 0.01 Ge$V^2$ $\leq$ b $\leq$ 0.40 Ge$V^2$
 \\
 0.05 GeV $\leq$ $\sigma$ $\leq$ 1.50 GeV
 \\
 4.00 GeV $\leq$ $m_{b}$ $\leq$ 5.00 GeV
 \\
 0.3 GeV $\leq$ $m_{q}$ $\leq$ 0.350 GeV\\ }

From the above range, the fitted parameters are tabulated in Table 1 to obtain the mass-spectra of mesons, diquarks and tetraquarks.  The mass-spectra of bottomonium mesons ($b\bar{b}$) and bottom (anti)diquark have been obtained from data set I. Similarly, the mass-spectra of B-mesons ($b\bar{q}$) and heavy-light (anti)diquark obtained from data set II. The mass-spectra of tetraquarks are computed using data set III. Indeed the mass-spectra of all bottom tetraquarks ($T_{4b}$) are calculated using data set I as well data set III. 

\begin{table}
\caption{The fitting parameters for obtaining the mass spectra of $bb\bar{b}\bar{b}$ and $bq\bar{b}\bar{q}$. The quark masses  $m_{b}$ = 4.783 GeV, and $m_{q}$ = 0.323 GeV have been taken from PDG \cite{pdg}.}
\label{tab:1}       
\begin{tabular}{llllllllllll}
\hline\noalign{\smallskip}
Data Set& $\alpha_{s}$ & $\sigma$ (GeV) & b (Ge$V^{2}$) \\
\noalign{\smallskip}\hline\noalign{\smallskip}
I& 0.3841 &1.50 & 0.1708 \\
II& 0.70 & 0.3049 & 0.0946  \\
III&   0.3714  &   1.50   &   0.1445 \\
\noalign{\smallskip}\hline
\end{tabular}
\end{table}

\begin{table*}
\centering
\caption{The mass-spectra of (1S-5S, 1P-3P and 1D-wave) bottomonium states [$b\bar{b}$], obtained from Data Set I. ($m_{i}^{exp}$) corresponds to recent PDG \cite{pdg}.}
\footnotesize
\begin{tabular*}{180mm}{@{\extracolsep{\fill}}cccccccccccccc}
\hline
$N^{2S+1}{L}_{J}$&$J^{PC}$	&	$\langle E\rangle$	&	$\langle V_{V}\rangle$	&	$\langle V_{S}\rangle$	&	$\langle V_{SS}\rangle$	&	$\langle V_{LS}\rangle$	&	$\langle V_{T}\rangle$ &	$\langle V^{(1)}(r_{ij}) \rangle $	&	$\langle K.E. \rangle$	&	$m_{i}^{f}$ 	& $m_{i}^{exp}$ & Meson \\
\hline
$1^{1}S_{0}$	&$0^{-+}$&	-133.0	&	-757.8	&	165.5	&	-23.4	&	0	&	0	&	-2.2	&	482.3	&	9409	&	9398.7$\pm2.0$&$\eta_{b}(1S)$\\
$1^{3}S_{1}$&$1^{--}$	&	-132.9	&	-757.8	&	165.5	&	7.6	&	0	&	0	&	-2.3	&	451.6	&	9440	&	9460.30$\pm0.26$&	$\Upsilon (1S)$\\
$2^{1}S_{0}$&$0^{-+}$	&	428.3	&	-352.1	&	404.5	&	-6.5	&	0	&	0	&	-1.1	&	382.5	&	9987	&...&...\\
$2^{3}S_{1}$&$1^{--}$	&	429.6	&	-352.1	&	404.5	&	2.1	&	0	&	0	&	-0.9	&	374.4	&	9997	&		10023.26$\pm0.31$&$\Upsilon (2S)$\\
$3^{1}S_{0}$&$0^{-+}$	&	764.1	&	-258.2	&	596.3	&	-4.2	&	0	&	0	&	-0.7	&	430.4	&	10325	&...&...\\
$3^{3}S_{1}$	&$1^{--}$&	764.1	&	-258.2	&	596.3	&	1.4	&	0	&	0	&	-0.7	&	425.1	&	10331	&	10355.2$\pm0.5$&	$\Upsilon (3S)$\\
$4^{1}S_{0}$&$0^{-+}$	&	1034.1	&	-213.1	&	761.0	&	-3.2	&	0	&	0	&	-0.5	&	489.5	&	10596	&...&...\\
$4^{3}S_{1}$&$1^{--}$	&	1034.1	&	-213.1	&	761.0	&	1.0	&	0	&	0	&	-0.5	&	485.4	&	10601&	10579.4$\pm1.2$	&	$\Upsilon (4S)$\\
$5^{1}S_{0}$&$0^{-+}$	&	1270.4	&	-185.4	&	909.1	&	-2.7	&	0	&	0	&	-0.4	&	549.4	&	10833	&...&...\\
$5^{3}S_{1}$&$1^{--}$	&	1270.4	&	-185.4	&	909.1	&	0.9	&	0	&	0	&	-0.4	&	546.1	&	10837	&...&...\\
$1^{3}P_{0}$	&$0^{++}$&	335.3	&	-313.6	&	329.1	&	0.8	&	-28.0	&	-10.8	&	-1.2	&	319.2	&	9863	&	9859.44$\pm0.42$&	$\chi_{b0} (1P)$\\
$1^{3}P_{1}$&$1^{++}$	&	335.3	&	-313.6	&	329.1	&	0.8	&	-14.0	&	5.4	&	-1.2	&	319.2	&	9893	&	9892.78$\pm0.26$&	$\chi_{b1} (1P)$\\
$1^{1}P_{1}$	&$1^{+-}$&	335.3	&	-313.6	&	329.1	&	-2.5	&	0	&	0	&	-1.2	&	322.7	&	9898	&	9899.3$\pm0.8$&	$h_{b} (1P)$\\
$1^{3}P_{2}$&$2^{++}$	&	335.3	&	-313.6	&	329.1	&	0.8	&	14.0	&	-1.0	&	-1.2	&	319.2	&	9915	&	9912.21$\pm0.26$&	$\chi_{b2} (1P)$\\
$2^{3}P_{0}$&$0^{++}$	&	679.2	&	-227.2	&	529.5	&	0.7	&	-23.3	&	-8.7	&	-0.7	&	376.4	&	10213	&	10232.5$\pm0.4$&	$\chi_{b0} (2P)$\\
$2^{3}P_{1}$&$1^{++}$	&	679.2	&	-227.2	&	529.5	&	0.7	&	-11.6	&	-8.7	&	-0.7	&	377.1	&	10239	&	10255.46$\pm0.22$&	$\chi_{b1} (2P)$\\
$2^{1}P_{1}$	&$1^{+-}$&	679.2	&	-227.2	&	529.5	&	-2.3	&	0	&	0	&	-0.7	&	380.2	&	10243&-	&	-\\
$2^{3}P_{2}$	&$2^{++}$&	679.2	&	-227.2	&	529.5	&	0.7	&	11.6	&	-0.8	&	-0.7	&	377.1	&	10257	&	10268.65$\pm0.22$&	$\chi_{b2} (2P)$\\
$3^{3}P_{0}$&$0^{++}$	&	955.3	&	-187.1	&	699.7	&	0.6	&	-21.4	&	-8.0	&	-0.5	&	442.4	&	10492	&...&...\\
$3^{3}P_{1}$&$1^{++}$	&	955.3	&	-187.1	&	699.7	&	0.6	&	-10.7	&	4.0	&	-0.5	&	442.4	&	10515	&	10513.42$\pm0.41$&	$\chi_{b1} (3P)$\\
$3^{1}P_{1}$	&$1^{+-}$&	955.3	&	-187.1	&	699.7	&	-2.0	&	0	&	0	&	-0.5	&	445.4	&	10519&...	&	...\\
$3^{3}P_{2}$&$2^{++}$	&	955.3	&	-187.1	&	699.7	&	0.6	&	10.7	&	-0.8	&	-0.5	&	442.4	&	10532	&10524.02$\pm0.57$&	$\chi_{b2} (3P)$\\
$1^{3}D_{1}$&$1^{--}$	&	576.4	&	-213.0	&	455.2	&	0.09	&	-6.2	&	-1.2	&	-0.9	&	333.7	&	10135&-	&	-\\
$1^{3}D_{2}$	&$2^{--}$&	576.4	&	-213.0	&	455.2	&	0.09	&	-2.0	&	1.2	&	-0.9	&	333.7	&	10141	&10163.7$\pm1.4$&	$\Upsilon_{2} (1D)$\\
$1^{1}D_{2}$	&$2^{-+}$&	576.4	&	-213.0	&	455.2	&	-0.2	&	0	&	0	&	-0.9	&	333.7	&	10142	&-&	-\\
$1^{3}D_{3}$&$3^{--}$	&	576.4	&	-213.0	&	455.2	&	0.09	&	-4.2	&	-0.3	&	-0.9	&	333.7	&	10146&-	&	-\\
[1ex]
\hline 
\end{tabular*}
\end{table*}

\begin{table*}
\centering
\caption{The mass-spectra of (1S-3S, 1P-3P and 1D-wave) B-mesons [$b\bar{q}$], obtained from Data Set II. ($m_{i}^{exp}$) corresponds to recent PDG \cite{pdg}.}
\footnotesize
\begin{tabular*}{180mm}{@{\extracolsep{\fill}}ccccccccccccc}
\hline
$N^{2S+1}L_{J}$&$J^{PC}$	&	$\langle E\rangle$	&	$\langle V_{V}\rangle$	&	$\langle V_{S}\rangle$	&	$\langle V_{SS}\rangle$	&	$\langle V_{LS}\rangle$	&	$\langle V_{T}\rangle$ &	$\langle V^{(1)}(r_{ij}) \rangle $	&	$\langle K.E. \rangle$	&	$m_{i}^{f}$ 	&	$m_{i}^{exp}$& Meson\\
\hline
$1^{1}S_{0}$&$0^{-+}$	&	207.3	&	-427.8	&	286.0	&	-9.2	&	0	&	0	&	-7.1	&	358.4	&	5304	&5279.65$\pm0.12$&	$B^{\pm}$\\
$1^{3}S_{1}$	&$1^{--}$&	207.3	&	-427.8	&	286.0	&	3.0	&0	&	0	&	-6.9	&	346.6	&	5317 &5324.70$\pm0.21$	&	$B^{*}$\\
$2^{1}S_{0}$&$0^{-+}$	&	774.4	&	-241.3	&	599.6	&	-3.0	&	0	&	0	&	-4.3	&	416.5	&	5877&-	&	-\\
$2^{3}S_{1}$&$1^{--}$	&	771.4	&	-241.3	&	599.6	&	1.0	&	0	&	0	&	-4.0&	415.0	&	5881&-	&	-\\
$3^{1}S_{0}$	&$0^{-+}$&	1173.2	&	-185.4	&	846.3	&	-2.0	&	0	&	0	&	-2.9	&	513.9	&	6277&-	&	-\\
$3^{3}S_{1}$&$1^{--}$	&	1173.2	&	-185.4	&	846.3	&	0.7	&	0	&	0	&	-2.6	&	512.3	&	6281&-	&	-\\
$1^{3}P_{0}$&$0^{++}$	&	626.3	&	-210.5	&	489.1	&	1.1	&	-28.0	&	-14.1	&	-4.7	&	346.4	&	5691	&-&	-\\
$1^{3}P_{1}$&$1^{++}$	&	623.7	&	-210.5	&	489.1	&	1.1	&	-14.0	&	6.9	&	-4.7	&	344.2	&	5723&5725.9$_{-2.7}^{+2.5}$	&	$B_{1}(5721)^{+}$\\
$1^{1}P_{1}$&$1^{+-}$&	623.7	&	-210.5	&	489.1	&	-3.3	&	0	&	0	&	-4.7	&	348.7	&	5726&5726.1$\pm1.3$	&	$B_{1}(5721)^{0}$\\
$1^{3}P_{2}$	&$2^{++}$&	623.7	&	-210.5	&	489.1	&	1.1	&	14.0	&	-1.0	&	-4.7	&	344.1	&	5743& 5739.5$\pm0.7$	&	$B^{*}_{2}(5747)^{0}$\\
$2^{3}P_{0}$&$0^{++}$	&	1040.0	&	-160.2	&	750.4	&	0.7	&	-27.3	&	-12.6	&	-3.4	&	449.3	&	6107&-	&	-\\
$2^{3}P_{1}$	&$1^{++}$&	1042.1	&	-160.2	&	750.4	&	0.7	&	-13.6	&	-6.3	&	-3.4	&	451.2	&	6141&-	&	-\\
$2^{1}P_{1}$&$1^{+-}$	&	1042.1	&	-160.2	&	750.4	&	-2.1	&	0	&	0	&	-3.4	&	454.2	&	6146&-	&	-\\
$2^{3}P_{2}$&$2^{++}$	&	1042.1	&	-160.2	&	750.4	&	0.7	&	13.6	&	-1.2	&	-3.4	&	451.3	&	6161&-	&	-\\
$3^{3}P_{0}$&$0^{++}$	&	1391.2	&	-134.5	&	974.0	&	0.5	&	-27.1	&	-12	&	-2.4	&551.3&	6458&-	&	-\\
$3^{3}P_{1}$&$1^{++}$	&	1391.2	&	-134.5	&	974.0	&	0.5	&	-13.5	&	-6.0	&	-2.4	&	551.2&6490&-	&	-\\
$3^{1}P_{1}$&$1^{+-}$	&	1391.2	&	-134.5	&	974.0	&	-1.6	&	0	&	0	&	-2.4	&553.2&	6495&-	&	-\\
$3^{3}P_{2}$	&$2^{++}$&	1391.2	&	-134.5	&	974.0	&	0.5	&	13.5	&	-1.2	&	-2.4	&551.2&	6510&-	&	-\\
$1^{3}D_{1}$&$1^{--}$	&	894.2	&	-150.5	&	649.2	&	0.3	&	-1.9	&	-1.8	&	-4.0	&	395.2	&	5996&-	&	-\\
$1^{3}D_{2}$	&$2^{--}$&	894.2	&	-150.5	&	649.2	&	0.3	&	-0.6	&	1.8	&	-4.0	&	395.2	&	6002&-	&	-\\
$1^{1}D_{2}$	&$2^{-+}$&	894.2	&	-150.5	&	649.2	&	-1.1	&	-1.9	&	-1.8	&	-4.0	&	397.0	&	5999&-	&	-\\
$1^{3}D_{3}$	&$3^{--}$&	894.2	&	-150.5	&	649.2	&	0.3	&	1.2	&	-0.5	&	-4.0	&	395.2	&	6001&-	&	-\\
[1ex]
\hline
\end{tabular*}
\end{table*}

\section{Results and Discussion}
\subsection{Bottomonium/ B-mesons}
\label{sec:2}
To calculate the mass-spectra of diquarks and tetra-quarks, first, we estimate the mass-spectra of bottomonium states [$b\bar{b}$] and B-mesons, whose results are tabulated in Table 2 and Table 3 , respectively. 
The model's reliability and consistency have been tested by obtaining the mass-spectra of heavy and heavy-light bottom mesons.
The SU(3) color symmetry allows only colorless quark combination $|Q\bar{Q}\rangle$ to form any color  singlet state \cite{griffiths,thesis}, as in our case [$b\bar{b}$] and [$b\bar{q}$] are mesons and exhibits $|Q\bar{Q}\rangle: \mathbf{3 \otimes\bar{3}=1\oplus 8}$ representation which leads to carry a color factor $k_{s}=-\frac{4}{3}$ \cite{debastiani}.
The masses of the particular [$b\bar{b}$] and [$b\bar{q}$] states are obtained namely

\begin{equation} 
M_{(b\bar{b})} = 2m_{b} + E_{b\bar{b}} + 
\langle V^{1}(r_{ij})\rangle
\end{equation}
and
\begin{equation} 
M_{(b\bar{q})} = m_{b}+ m_{\bar{q}}+ E_{b\bar{q}} + \langle V^{1}(r_{ij})\rangle
\end{equation}
The final masses obtained from the above expression constitute the contributions from different spin-dependent terms (spin-spin, spin-orbital and tensor) have tabulated in Table 2 and Table 3.
The mass-spectra of the mesons produced in this study are compatible with the experimental data available in the most recent updated PDG \cite{pdg}. Additionally, the current work's findings are consistent with those in Ref. \cite{prd}, where the author computed the mass-spectra of all-charm [$cc\bar{c}\bar{c}$], all-bottom [$bb\bar{b}\bar{b}$], and heavy-light tetraquarks [$Qq\bar{Q}\bar{q}$] (Q = b, c and q = u, d).
\\
\\
There are a total of 15 bottomonium mesons ($b\bar{b}$) produced from the model, all of which have masses fairly closed to those predicted experimentally. In the case of bottomonium S-wave states the discrepancy ($\Delta$) is around 30 MeV. Particularly, in vector states $1^{3}S_{1}$ where spin-spin interaction contributes repulsive strength which maximizes $\Delta$.
Similarly, there are B-mesons whose masses have also good agreement with experimentally predicted data and have discrepancy ($\Delta$) nearly 30 MeV. 
At high energy scale, discrepancy nearly 30 MeV's between the model's mass and experimental data can be tolerated and the fitting parameters are assumed as best fit. 
\\
\\
Spin-dependent interactions are crucial in heavy quarkonium study specially in bottomonium and charmonium because they allow for the consideration of QCD dynamics in the heavy quark scenario, which lies between the perturbative and non-perturbative regimes \cite{thesis}. The spin-spin interaction's involvement in orbitally excited states is particularly intriguing. We notice that the masses of P-wave states are more precise than those of S-wave states owing to spin-orbit and tensor contributions (see Table 2 and Table 3).
The slighter contribution of the relativistic term, which is greater for lighter quarks, alters the spectrum by a few MeV's.

\begin{table*}
\centering
\caption{The masses vector (anti)diquarks  and energy eigenvalue from present work and comparison with other prior works. Units are in (MeV).}
\label{tab:1}       
\begin{tabular}{lllllllll}
\hline
Diquark &$\langle E^{0}\rangle$ & Ours & \cite{prd} & \cite{mb}& \cite{mk,de}& \cite{mn,zg} &\cite{Su,RK}\\
\hline
 bb &78.3 & 9641 & 9643 & 9845 & 9718 & 9850 & 8670$\pm$690\\
 bq & 235.7 & 5331 & 5339& 5465 &5381&5130$\pm$110&5080$\pm$40\\
\noalign{\smallskip}\hline
\end{tabular}
\end{table*}
\subsubsection{Diquarks}
A (anti)diquark is a pair of (anti)quarks that interact with one another through gluonic exchange and can form a bound state (qq) \cite{diquarks}. It's important to remember that while constructing a diquark is a composite (qq) system not a point-like object. The form factor, which can be represented as the overlap integral of diquark wave functions, does indeed smear its interaction with gluons. The Pauli principle should also be considered, which results in the following ground state diquark limitations.
To comply the Pauli exclusion principle, which states that diquarks with the same flavour quark should have a spin of 1, the diquark's wavefunction must be antisymmetric \cite{griffiths}. The (qq') diquark, which is made of quarks of various flavours, may have spins S = 0,1 (scalar [qq'], axial vector $\lbrace qq'\rbrace$ diquarks), while the $\lbrace qq\rbrace$ diquark, which is composed of quarks of the same flavour, can only have spin S = 1. Because of the stronger attraction owing to the spin–spin interaction, the scalar S diquark is frequently referred to as a ``good" diquark, while the heavier axial vector diquark is referred to as a ``bad" diquark \cite{exotic2}.
\\
\\
To produce the most compact diquark, we will utilise the ground state ($1^{3}S_{1}$) axial vector diquarks $\lbrace bb\rbrace$ or $\lbrace bq\rbrace$, which have no orbital or radial excitations.
%
Everything done for quark-quark interactions is assumed to be identical to antiquark-antiquark interactions, with the exception that colors are replaced by anticolors. We utilised the same methodology as in the case of the bottom mesons to get the mass-spectra of (anti)diquarks.
According to QCD color symmetry, two quarks are combined in the fundamental ({\bf3}) representation to obtain the diquark, presented by  $\mathbf{3 \otimes 3 = \bar{3} \oplus 6}$.
Moreover, antiquarks are combined in the $\mathbf{\bar{3}}$ representation and can be presented as $\mathbf{\bar{3} \otimes \bar{3} = 3 \oplus \bar{6}}$ \cite{debastiani,griffiths}. 
\textbf{The diquark-antidiquark approximation is significant because it reduces a complex four-body problem to a simple two-body problem \cite{deutron,deutron1}. The hamiltonian, on the other hand, ceases replicating the meson spectra when doing the full four-body  basis treatment \cite{diquark1}. The explanation for this is simple: the $3 \otimes \bar{3}$ color coupling can be transformed into a $1\otimes1$ state, and also a $8\otimes8$ state.}
The QCD color symmetry  produces a color factor $k_{s}=-\frac{2}{3}$ in antitriplet state and makes the short distance part $(\frac{1}{r_{ij}})$ of the interaction attractive \cite{griffiths}.
\\
\\
We compared the diquark masses acquired in this work to those obtained in the other prior investigations mentioned in Table 4.
The diquark masses are investigated in Ref. \cite{mb} using the so-called Schwinger-Dyson and Bethe-Salpeter approach, which account for kinetic energy as well as splittings in the spin-spin, spin-orbit, and tensor interactions.
The masses of diquarks calculated in this research are consistent with \cite{prd} and are less than those reported in Ref. \cite{mb}.
Relativistic models, such as those presented in Refs. \cite{mk,de,mn}, all predict larger diquark masses, whereas models based on QCD sum rules, such as those presented in Refs. \cite{zg,Su,RK} all predict smaller diquark masses. The discrepancies may be due to the addition of new and updated data in this study.

\begin{table*}
\centering
\caption{ The mass-spectra of S and P-wave $T_{4b}$ tetraquark states, obtained from Data Set I whereas $m_{th}$ corresponds to mass of two meson threshold.} 
\footnotesize

\begin{tabular*}{180mm}{@{\extracolsep{\fill}}ccccccccccccc}
\hline
$N^{2S+1}{L}_{J}$&$J^{PC}$	&	$\langle E\rangle$	&	$\langle V_{V}\rangle$	&	$\langle V_{S}\rangle$	&	$\langle V_{SS}\rangle$	&	$\langle V_{LS}\rangle$	&	$\langle V_{T}\rangle$ &	$\langle V^{(1)}(r_{ij}) \rangle $	&	$\langle K.E. \rangle$	&	$m_{i}^{f}$  & $m_{th}$&Threshold\\
\hline
$1^{1}S_{0}$ &$0^{++}$	&	-532.8	&	-1344	&	95.4	&	-30.1	&	0	&	0	&	-2.7	&	745.3	&	18719	& 18798& $\eta_{b}(1S)\eta_{b}(1S)$\\
$1^{3}S_{1}$ &$1^{+-}$&	-532.8	&	-1344	&	95.4	&	-15.1	&	0	&	0	&	-2.7	&	730.4	&	18734&		18859& $\eta_{b}(1S)\Upsilon(1S)$\\
$1^{5}S_{2}$&$2^{++}$	&	-532.8	&	-1344	&	95.4	&	15.2	&	0	&	0	&	-2.7	&	701.0	&	18764	&	18920&$\Upsilon(1S)\Upsilon(1S)$\\
$2^{1}S_{0}$ &$0^{++}$	&	164.2	&	-589.5	&	277.1	&	-5.2	&	0	&	0	&	-1.0	&	481.5	&	19441		& 19998&$\eta_{b}(2S)\eta_{b}(2S)$\\
$2^{3}S_{1}$ &$1^{+-}$	&	164.2	&	-589.5	&	277.1	&	-2.6	&	0	&	0	&	-1.0	&	478.5	&	19443	& ...&...\\
$2^{5}S_{2}$ &$2^{++}$	&	164.2	&	-589.5	&	277.1	&	2.6	&	0	&	0	&	-1.0	&	473.1	&	19448	&	...&...\\
$3^{1}S_{0}$ &$0^{++}$	&	480.1	&	-427.4	&	435.3	&	-3.1	&	0	&	0	&	-0.6	&	475.3	&	19759	&	...&...\\
$3^{3}S_{1}$ &$1^{+-}$	&	480.1	&	-427.4	&	435.3	&	-1.5	&	0	&	0	&	-0.6	&	473.7	&	19760		&...&...\\
$3^{5}S_{2}$ &$2^{++}$	&	480.1	&	-427.4	&	435.3	&	1.5	&	0	&	0	&	-0.6	&	470.6	&	19764	&...&...\\
$1^{1}P_{1}$	&$1^{--}$&	105.1	&	-464.4	&	225.5	&	-5.8	&	0	&	0	&	-1.2	&	350.0	&	19381	&...&...\\
$1^{3}P_{0}$ &$0^{-+}$	&	105.1	&	-464.4	&	225.5	&	-2.9	&	-25.6	&	-18.1	&	-1.2	&	391.0	&	19340&19258	&$\eta_{b}(1S)\chi_{b0} (1P)$\\
$1^{3}P_{1}$&$1^{-+}$ 	&	105.1	&	-464.4	&	225.5	&	-2.9	&	-12.8	&	9.0	&	-1.2	&	350.7	&	19380&19292		&$\eta_{b}(1S)\chi_{b1} (1P)$	\\
$1^{3}P_{2}$ &$2^{-+}$	&	105.1	&	-464.4	&	225.5	&	-2.9	&	12.8	&	-1.8	&	-1.2	&	336.0	&	19395&19311		&	$\eta_{b}(1S)\chi_{b2} (1P)$ \\
$1^{5}P_{1}$ &$1^{--}$	&	105.1	&	-464.4	&	225.5	&	2.9	&	-38.4	&	-12.7	&	-1.2	&	392.0	&	19338&19298	&$\eta_{b}(1S)h_{b}(1P)$ \\
$1^{5}P_{2}$&$2^{--}$ 	&	105.1	&	-464.4	&	225.5	&	2.9	&	-12.8	&	12.7	&	-1.2	&	341.0	&	19390	&19353&$\Upsilon(1S)\chi_{b1} (1P)$\\
$1^{5}P_{3}$&$3^{--}$	&	105.1	&	-464.4	&	225.5	&	2.9	&	25.6	&	3.6	&	-1.2	&	319.0	&	19412&19372		&$\Upsilon(1S)\chi_{b2} (1P)$ \\
\hline
\end{tabular*}
\end{table*}

\begin{table*}
\centering
\caption{ The mass-spectra of S and P-wave tetraquark states $T_{4b}$, obtained from Data Set I+III whereas $m_{th}$ corresponds to mass of two meson threshold.} 
\footnotesize
\begin{tabular*}{180mm}{@{\extracolsep{\fill}}ccccccccccccc}
\hline
$N^{2S+1}{L}_{J}$&$J^{PC}$	&	$\langle E\rangle$	&	$\langle V_{V}\rangle$	&	$\langle V_{S}\rangle$	&	$\langle V_{SS}\rangle$	&	$\langle V_{LS}\rangle$	&	$\langle V_{T}\rangle$ &	$\langle V^{(1)}(r_{ij}) \rangle $	&	$\langle K.E. \rangle$	& $m_{i}^{f}$ &  $m_{th}$ & Threshold\\
\hline

$1^{1}S_{0}$&$0^{++}$	&	-503.4	&	-1252.3	&	83.4	&	-28.3	&	0	&	0	&	-2.3	&	693.2	&	18749&	18798& $\eta_{b}(1S)\eta_{b}(1S)$	 \\
$1^{3}S_{1}$&$1^{+-}$	&	-503.4	&	-1252.3	&	83.4	&	-14.1	&	0	&	0	&	-2.0	&	679.3	&	18764	&18859& $\eta_{b}(1S)\Upsilon(1S)$ \\
$1^{5}S_{2}$&$2^{++}$	&	-503.4	&	-1252.3	&	83.4	&	14.0	&	0	&	0	&	-2.3	&	651.2	&	18792	&18920&$\Upsilon(1S)\Upsilon(1S)$	 \\
$2^{1}S_{0}$	&$0^{++}$ &	136.1	&	-531.0	&	245.1	&	-4.7	&	0	&	0	&	-0.8	&	427.5	&	19414	&	19998&$\eta_{b}(2S)\eta_{b}(2S)$\\
$2^{3}S_{1}$ &$1^{+-}$&	136.1	&	-531.0	&	245.1	&	-2.3	&	0	&	0	&	-0.8	&	425.6	&	19416	&	...&... \\
$2^{5}S_{2}$&$2^{++}$	&	136.1	&	-531.0	&	245.1	&	2.3	&	0	&	0	&	-0.8	&	420.2	&	19421	&	...&... \\
$3^{1}S_{0}$&$0^{++}$	&	422.4	&	-379.4	&	387.3	&	-2.8	&	0	&	0	&	-0.5	&	417.5	&	19701	&	...&... \\
$3^{3}S_{1}$&$1^{+-}$	&	422.4	&	-379.4	&	387.3	&	-1.4	&	0	&	0	&	-0.5	&	416.5	&	19703	&	...&... \\
$3^{5}S_{2}$	&$2^{++}$&	422.4	&	-379.4	&	387.3	&	1.4	&	0	&	0	&	-0.5	&	417.5	&	19706	&	...&... \\
$1^{1}P_{1}$&$1^{--}$	&	84.2	&	-428.2	&	200.0	&	-5.0	&	0	&	0	&	-0.9	&	318.0	&	19361	&...&...	 \\
$1^{3}P_{0}$	&$0^{-+}$ &	84.2	&	-428.2	&	200.0	&	-2.5	&	-21.5	&	-15.2	&	-0.9	&	316.2	&	19327	&	 19258	&$\eta_{b}(1S)\chi_{b0} (1P)$\\
$1^{3}P_{1}$&$1^{-+}$	&	84.2	&	-428.2	&	200.0	&	-2.5	&	-10.4	&	7.7	&	-0.9	&	316.0	&	19361	&	19292		&$\eta_{b}(1S)\chi_{b1} (1P)$ \\
$1^{3}P_{2}$&$2^{-+}$	&	84.2	&	-428.2	&	200.0	&	-2.5	&	10.5	&	-1.5	&	-0.9	&	316.0	&	19373	&19311		&	$\eta_{b}(1S)\chi_{b2} (1P)$ 	 \\
$1^{5}P_{1}$&$1^{--}$	&	84.2	&	-428.2	&	200.0	&	2.5	&	-32.4	&	-11.1	&	-0.9	&	311.0	&	19325	&19298	&$\eta_{b}(1S)h_{b}(1P)$	 \\
$1^{5}P_{2}$&$2^{--}$	&	84.2	&	-428.2	&	200.0	&	2.5	&	-11.2	&	11.0	&	-0.9	&	311.2	&	19369	&	19353&$\Upsilon(1S)\chi_{b1} (1P)$ \\
$1^{5}P_{3}$&$3^{--}$	&	84.2	&	-428.2	&	200.0	&	2.5	&	21.7	&	-3.0	&	-0.9	&	311.0	&	19388	&	19372		&$\Upsilon(1S)\chi_{b2} (1P)$\\
\hline
\end{tabular*}
\end{table*}

\begin{table*}
\centering
\caption{ The mass-spectra of S and P-wave $bq\bar{b}\bar{q}$ tetraquark states, obtained from Data Set II+III, whereas $m_{th}$ corresponds to mass of two meson threshold.}  
\footnotesize
\begin{tabular*}{170mm}{@{\extracolsep{\fill}}ccccccccccccc}
\hline
$N^{2S+1}{L}_{J}$ 	&$J^{PC}$	&	$\langle E\rangle$	&	$\langle V^{(0)}_{V}\rangle$	&	$\langle V^{(0)}_{S}\rangle$	&	$\langle V^{(1)}_{SS}\rangle$	&	$\langle V^{(1)}_{LS}\rangle$	&	$\langle V^{(1)}_{T}\rangle$	& $V^{(1)}(r)$	&	$m_{i}^{f}$ &  $m_{th}$ & Threshold   \\
\hline
$1^{1}S_{0}$	&$0^{++}$&	-181.1	&	-764.0	&	135.0	&	-50.0	&	0	&	0	&	-2.0	&	10429	&10558&	$B^{\pm}B^{\pm}$\\
$1^{3}S_{1}$	&$1^{+-}$&	-181.1	&	-764.0	&	135.0	&	-25.5	&	0	&	0	&	-2.0	&10454	&10603 &$B^{\pm}B^{*}$	 \\
$1^{5}S_{2}$&$2^{++}$	&	-181.1	&	-764.0	&	135.0	&	25.4	&	0	&	0	&	-2.0	&	10505	&	10648 & $B^{*}B^{*}$ \\
$2^{1}S_{0}$&$0^{++}$	&	339.1	&	-340.0	&	341.0	&	-13.3	&	0	&	0	&	-0.9	&	10987&...&...	 \\
$2^{3}S_{1}$&$1^{+-}$	&	339.1	&	-340.0	&	341.0	&	-6.6	&	0	&	0	&	-0.8	&	10995	&...&...	 \\
$2^{5}S_{2}$&$2^{++}$	&	339.1	&	-340.0	&	341.0	&	6.6	&	0	&	0	&	-0.9	&	11009&...&...	 \\
$3^{1}S_{0}$&$0^{++}$	&	637.4	&	-247.2	&	507.0	&	-8.4	&	0	&	0	&	-0.6	&	11290	&...&...	 \\
$3^{3}S_{1}$&$1^{+-}$	&	637.4	&	-247.2	&	507.0	&	-4.2	&	0	&	0	&	-0.6	&	11295	&...&...	 \\
$3^{5}S_{2}$&$2^{++}$	&	637.4	&	-247.2	&	507.0	&	4.2	&	0	&	0	&	-0.6	&	11303	&...&...	 \\
$1^{1}P_{1}$&$1^{--}$	&	262.0	&	-305.0	&	277.0	&	-5.5	&	0	&	0	&	-0.8	&	10918	&	...&... \\
$1^{3}P_{0}$	&$0^{-+}$&	262.0	&	-305.0	&	277.0	&	-2.7	&	-23.0	&	-17.4	&	-0.8	&	10881	&11006	&$B^{\pm}B_{1}(5721)^{+}$ \\
$1^{3}P_{1}$&$1^{-+}$	&	262.0	&	-305.0	&	277.0	&	-2.7	&	-11.5	&	8.7	&	-0.8	& 10918	&...&... \\
$1^{3}P_{2}$&$2^{-+}$	&	262.0	&	-305.0	&	277.0	&	-2.7	&	11.5	&	1.7	&	-1.0	&	10931	&...&... \\
$1^{5}P_{1}$&$1^{--}$	&	262.0	&	-305.0	&	277.0	&	2.7	&	-34.4	&	-12.2	&	-0.8	&	10880&...&... \\
$1^{5}P_{2}$&$2^{--}$	&	262.0	&	-305.0	&	277.0	&	2.7	&	-11.5	&	-12.2	&	-1.0	&	10928	&...&...	 \\
$1^{5}P_{3}$&$3^{--}$	&	262.0	&	-305.0	&	277.0	&	-5.5	&	23.0	&	-3.4	&	-0.8	&	10946	&...&...	 \\
\hline
\end{tabular*}
\end{table*}

\begin{table*}
\centering
\caption{Comparison of the tetraquarks masses from the present work with others}
\begin{tabular}{lllllllllllll}
\hline\noalign{\smallskip}
$bb\bar{b}\bar{b}$ State & $J^{PC}$  & Ours & \cite{prd}&\cite{a.v.}& \cite{ming}&\cite{kar} & \cite{guang}& \cite{deng}& \cite{qi}& \cite{xin}&\cite{ma}&\cite{Xi}\\
\noalign{\smallskip}\hline\noalign{\smallskip}
$1^{1}S_{0}$ &$0^{++}$ & 18719&18723&18754&19322 &18826$\pm$25&...&19329&19255& 19122-19344&18748&19178 \\
$1^{3}S_{1}$&$1^{+-}$&18734&18738&18808& 19329&  ..... &19247&19373&19251&19184-19354&18828&19226\\
$1^{5}S_{2}$&$2^{++}$&18764&18768&18916&19341&18956$\pm$25&19249&19387&19262&19236-19374&18900 &19236\\
$1^{1}P_{1}$&$1^{--}$&19361&...&...&...&...&...&...&...&...&19281&...\\
$1^{3}P_{0}$&$0^{-+}$&19327&...&...&...&...&...&...&...&...&19288&...\\
$1^{3}P_{1}$&$1^{-+}$&19361&...&...&...&...&...&...&...&...&...&...\\
$1^{3}P_{2}$&$2^{-+}$&19373&...&...&...&...&...&...&...&...&...&...\\
$1^{5}P_{1}$&$1^{--}$&19325&...&...&...&...&...&...&...&...&...&...\\
$1^{5}P_{2}$&$2^{--}$&19369&...&...&...&...&...&...&...&...&...&...\\
$1^{5}P_{3}$&$3^{--}$&19388&...&...&...&...&...&...&...&...&...&...\\
\\
\hline
$bq\bar{b}\bar{q}$ State& $J^{PC}$  & Ours&\cite{aslam} & \cite{prd}&\cite{de}& \cite{mh}&\cite{farhad} &\cite{ebert}&\cite{smpatel}\\
\hline
$1^{1}S_{0}$ & $0^{++}$ & 10429&...&10445&10473&10469&10462&10471&10143 \\

$1^{3}S_{1}$&$1^{+-}$&10454&...&10472&10494&10453&...&10492&10233\\
$1^{5}S_{2}$&$2^{++}$&10505&10520&10523&10534&....&...&10534&10413\\
$1^{1}P_{1}$&$1^{--}$&10918&10900&10936&10807&...&10819&10807&...\\
$1^{3}P_{0}$&$0^{-+}$&10881&...&...&...&...&...&10836&...\\
$1^{3}P_{1}$&$1^{-+}$&10918&...&...&10939&...&...&10847&...\\
$1^{3}P_{2}$&$2^{-+}$&10931&...&...&...&...&...&10854&...\\
$1^{5}P_{1}$&$1^{--}$&10880&...&...&10944&...&...&10827&...\\
$1^{5}P_{2}$&$2^{--}$&10928&...&...&...&...&...&10856&...\\
$1^{5}P_{3}$&$3^{--}$&10946&...&...&...&...&...&10858&...\\
\noalign{\smallskip}\hline
\end{tabular}
\end{table*}


%
\subsection{Tetraquark}
Tetraquarks are color singlet states made up of a diquark and an antidiquark in color antitriplet $\mathbf{\bar{3}}$ and triplet $\bf{3}$ configurations respectively, that are held together by color forces \cite{de}. The four-body non-relativistic calculation is simplified to a  two-body calculation using this approximation \cite{debastiani,prd,de}. A $T_{4b}$ and $bq\bar{b}\bar{q}$ are color singlet states and yield a color factor $k_{s} = -\frac{4}{3}$.
The $1^3S_{1}$ diquark-antidiquark are combined to form color singlet tetraquark \cite{thesis}, and that can be represented as; $|QQ|^{3} \otimes|\bar{Q}\bar{Q}|^{\bar{3}}\rangle = \mathbf{1  \oplus 8}$. The mass-spectra of all bottom ($T_{4b}$) and heavy-light bottom tetraquarks ($bq\bar{b}\bar{q}$) have been obtained with the same formulation as in the case of mesons, namely;

\begin{equation}
M_{(T_{4b})} = m_{bb}+ m_{\bar{b}\bar{b}} + E_{[bb][\bar{b}\bar{b}]} + \langle V^{1}(r_{ij})\rangle
\end{equation}
and
\begin{equation}
M_{bq\bar{b}\bar{q}} = m_{bq}+ m_{\bar{b}\bar{q}} + E_{[bq][\bar{b}\bar{q}]} + \langle V^{1}(r_{ij})\rangle
\end{equation}
The masses of $T_{4b}$ and $bq\bar{b}\bar{q}$ arise mostly from the cornell potential and the relativistic correction term $\langle V^{1}(r)\rangle$. All spin-dependent terms have been computed for spin-1 diquarks and antiquarks that combine to produce a color singlet tetraquark with spin $S_{T}$ = 0,1,2. The mass-spectra of radial and orbital excitations are obtained by coupling the total spin $S_{T}$ with the total orbital angular momentum $L_{T}$, which results in the total angular momentum $J_{T}$.
To get the tetraquark's total spin $S_{T}$ and quantum number $J^{PC}$, we shall utilise notations for the spins of a diquark $S_{d}$ and an antidiquark $S_{\bar{d}}$. The interaction of $S_{T}$ with the orbital angular momentum $L_{T}$ results in the formation of a color singlet state $S_{T} \otimes L_{T}$.
\begin{equation}
|T_{4Q}\rangle = |S_{d},S_{\bar{d}},S_{T},L_{T} \rangle_{J_{T}}
\end{equation}
To find out the quantum numbers ($J^{PC}$) of the tetra-quark states, one can use the following formula; $P_{T}=(-1)^{L_{T}}$ and $C_{T}=(-1)^{L_{T} + S_{T}}$.
\\
\\
The masses of low-lying S-wave $T_{4b}$ and $bq\bar{b}\bar{q}$ states are anticipated to be in the range of 18-20 GeV and 10-11 GeV, respectively \cite{jw}, in the current study, the masses are also found to be in this range.
As shown in Table 5, Table 6, and Table 7, the compactness of the 1S-states is mostly due to the coulomb interaction. This indicates that one-gluon exchange is the dominant mechanism behind the strong interaction between diquarks and antidiquarks, which results in a negative energy eigenvalue E. The contribution of the confinement term increases with the increase in radial and orbital states.
\\
\\
Within the specific tetraquark mass-spectrum, the attractive strength of the spin-spin interaction decreases as the number of radial and orbital excited states increases. In this instance, we must bear in mind that the factors originating from $S_{1}$ and $S_{2}$ are greater for the coupling of two spin-1 particles than for the coupling of two spin-$\frac{1}{2}$ particles. It is worth noting that, despite the fact that the spin-dependent terms have been suppressed by a factor $\frac{1}{m_{qq}^{2}}$, one would anticipate them to be less than the equivalent terms in $q\bar{q}$ mesons. The color interaction brings diquark and antidiquark so close together that the suppression caused by this component $\frac{1}{m_{qq}^{2}}$, is swamped by the massive suppression at the system's origin. On the other hand, spin-orbital and the tensor interactions becomes more important when radial or orbital excitations are involved. When the relativistic term $V^{1}(r_{ij})$ is incorporated in the central potential, the mass spectrum shifts by a few MeVs. 
\\
The masses of 1S-wave all-bottom tetraquarks ($T_{4b}$) obtained from data Sets I and I+III are 100-200 MeV below the two-meson threshold, implying that these states may be accounted by the two meson thresholds stated in Tables 5 and 6. Initial predictions for $bb\bar{b}\bar{b}$ below the 2$\Upsilon$ threshold were made in Ref. \cite{first}, and these were confirmed by subsequent works \cite{nre,qcdsum,jw}.
\\
\\
\textbf{In Ref. \cite{Ciaran Hughes}, the authors used a first-principles lattice non relativistic QCD method to examine a $bb\bar{b}\bar{b}$ tetraquark bound state and found no evidence for the mass below the lowest non-interacting bottomonium-pair threshold in the $0^{++}$, $1^{+-}$, and $2^{++}$ channels.
Additionally, in Ref. \cite{silver}, the author employs a model with only chromomagnetic interactions and finds an unbound tetraquark and the same conclusion was drawn within non-relativistic \cite{nonrelativistic} and semi-relativistic quark-model studies \cite{relativistic}. The existence of all-bottom four-quark states is not ruled out by these result, but it draw attention to the gap between the diquark approximation and a completes treatment of the color basis. The author of Ref. \cite{a.v.}, finds that $0^{++}$, $1^{+-}$, and $2^{++}$ tetraquarks are confined by 44, 51, and 5 MeV, respectively, utilising a phenomenologically motivated non confining potential between the pointlike diquark and antidiquark. A tetraquark candidate has been identified about 300 MeV below the experimental 2$\eta_{b}$ threshold in the QCD sum-rules framework \cite{recent1}.}
Gang Yang et al., in Ref. \cite{Cbecchi}, predicted that the mass of the $0^{++}$ tetraquark $bb\bar{b}\bar{b}$ state would be lower than the 2$\eta_{b}$ threshold whereas in Ref. \cite{qcdsum}, a tetraquark candidate has been found between the experimental 2$\eta_{b}$ and 2$\Upsilon$ thresholds.
\\
\\
\\
Other S-wave candidates include $\eta_{b}\Upsilon$ and 2$\Upsilon$ for $1^{+-}$ and $2^{++}$, respectively, which have a mass discrepancies $\Delta$ of less than 200 MeV with a two-meson threshold. The lowest and greatest discrepancies for 1S-wave between the model's mass and the two-meson threshold are 49 MeV and 156 MeV, respectively, for data sets I and III. The discrepancy ($\Delta$) is lower in the mass-spectra obtained by data set III as compared to data set I which shows the fitting parameters from data set III are accurately fitted. Thus, the authors find that the $T_{4b}$ tetraquark state may be either below or between the 2$\eta_{b}$ and 2$\Upsilon$ thresholds (where the thresholds were established using experimental meson masses in both instances). 
\\
\\
Similarly, the P-wave masses match fairly well to the model's mass and two-meson threshold. As shown in Table 8, due to the scarcity of orbitally excited data, this research, in conjunction with others, seems to be helpful for studying orbitally excited states. In the P-wave states, where the contribution of spin-orbit and tensor terms is produced, the difference ($\Delta$) between two-meson thresholds and the model's mass falls to less than 100 MeV. The two meson $\eta_{b}\chi_{b1}$ threshold, which contains the quantum number ($1^{-+}$), and the $\Upsilon\chi_{b2}$ threshold, which contains the quantum number ($3^{--}$), have the greatest and smallest deviations from the two-meson threshold, respectively, of 88 MeV and 16 MeV. 
%
\\
\\
The masses of S-wave heavy-light bottom tetraquark states $bq\bar{b}\bar{q}$, are in good agreement with $B^{\pm}B^{\pm}$, $B^{\pm}B^{*}$, and $B^{*}B^{*}$ meson thresholds, with a difference of less than 200 MeV between the two meson thresholds ($m_{th}$) and the model's mass ($m^{f}_{i}$). The two most discussed bottom resonances, $Z_{b}(10610)$ and $Z_{b}(10650)$, both with ($1^{+-}$), may be recognised as possible candidates for $bq\bar{b}\bar{q}$ states \cite{zb2}, which have a mass variation of 150 MeV from the model's mass ($m^{f}_{i}$).


\section{Conclusion}
In the present work we have calculated the mass-spectra of all-bottom $[bb][\bar{b}\bar{b}]$ and heavy-light bottom $[bq][\bar{b}\bar{q}]$ tetraquarks, in a non-relativistic framework that includes the cornell like potential along with the relativistic correction term to the potential and spin-dependent interactions. Tetraquarks have been considered to be comprised of axial-vector diquarks and antidiquarks in a color antitriplet-triplet ($\bar{3}_{c}-3_{c}$) configuration. We first estimated the masses of bottom mesons to fit the model's free parameters, and then calculated the masses of axial diquarks to get the mass spectra of corresponding tetraquarks without violating the Pauli exclusion principle. 
%
%
%
\\
\\
In this way we predict the masses of diquarks and tetra-quarks which have atleast one bottom quark in the substructure. The doubly hidden-bottom tetraquark states have a considerably greater energy than ordinary bottomonium mesons and they may be differentiated experimentally from ordinary $q\bar{q}$ states. We explored the two most discussed $Z_{b}(10610)$ and $Z_{b}(10650)$ states with quantum numbers ($1^{+-}$), which are predicted as tetra-quark \cite{aslam2} and as molecule \cite{zb2}.
In the present study, the mass of the tetraquark state $bq\bar{b}\bar{q}$ ($1^{+-}$) is about 150 MeV distant from the experimental mass of these $Z_{b}$ states. 
When compared to the $BB^{*}$ and $B^{*}B^{*}$ states, the $Z_{b}(10610)$ and $Z_{b}(10650)$ states are just a few MeV above the two-meson threshold, indicating that these states should have a tetraquark structure above the threshold. To get a definite conclusion about the substructure of these $Z_{b}'s$ states, one must investigate their decay properties, which will be the subject of future extension work to this study.
\\
\\
Due to the fact that fully bottom tetraquark states $bb\bar{b}\bar{b}$ are heavier than heavy-light tetraquark states ($bq\bar{b}\bar{q}$), and they are likely to be recognised below two meson thresholds, namely 2$\eta_{b}$, 2$\Upsilon$ or $\eta_{b}\Upsilon$ with masses ranging from 18.7 GeV to 19 GeV. Our findings are in excellent accord with other non-relativistic models and other studies cited in the literature \cite{prd,a.v.,ma,Xi,ebert}. 
\\
\\
The production of the ($QQ\bar{Q}\bar{Q}$), states is very challenging since it necessitates the creation of two heavy quark pairs. However, new observations of the $J/\psi\Upsilon$ \cite{psiups}, pair and simultaneous $\Upsilon\Upsilon$ \cite{cms} events have given some insight on how these doubly hidden-charm/bottom tetra-quarks are produced. As a result, the 2$\eta_{b}$, $\eta_{b}\Upsilon$ and 2$\Upsilon$ channels may be excellent candidates for searching for the doubly hidden-bottom states $bb\bar{b}\bar{b}$, and LHCb, D0, and CMS could be suitable platforms.

\section*{Acknowledgement}
Rohit Tiwari was inspired by the work of V.R. Debastiani, F.S. Navarra and Per Lundhammar, Tommy Ohlsson on Spectroscopic study in non-relativistic model for the tetraquark and would like to thank them for their valuable contributions to this field.




\end{document}